# Earliest Datable Records of Aurora-like Phenomena in the Astronomical Diaries from Babylonia

Author #1: Hisashi Hayakawa, Graduate School of Letters, Kyoto University, Kyoto, Japan, hayakawa@kwasan.kyoto-u.ac.jp

Author #2: Yasuyuki Mitsuma, Graduate School of Arts and Sciences, The University of Tokyo, Tokyo, Japan, licorne@soleil.ocn.ne.jp

Author #3: Yusuke Ebihara, Research Institute for Sustainable Humanosphere, Kyoto University; Unit of Synergetic Studies for Space, Kyoto University, Kyoto, Japan, ebihara@rish.kyoto-u.ac.jp

Author #4: Akito Davis Kawamura, Kwasan Observatory, Kyoto University, Kyoto, Japan, akitodk@kwasan.kyoto-u.ac.jp

Author #5: Hiroko Miyahara, Musashino Art University, Tokyo, Japan, miyahara@musabi.ac.jp

Author #6: Harufumi Tamazawa, Kwasan Observatory, Kyoto University, Kyoto, Japan, tamazawa@kwasan.kyoto-u.ac.jp

Author #7: Hiroaki Isobe, Unit of Synergetic Studies for Space, Kyoto University; Graduate School of Advanced Integrated Studies for Human Survivability, Kyoto University, Japan, isobe@kwasan.kyoto-u.ac.jp

**ABSTRACT**

The Astronomical Diaries from Babylonia (ADB) are an excellent source of information of natural phenomena, including astronomical ones, in pre-Christ era because it contains the record of highly continuous and systematic observations. In this article we present results of a survey of aurora-like phenomena in ADB, spanning from BCE 652 to BCE 61. We have found 9 records of aurora-like phenomena. Philological and scientific examinations suggest 5 of them can be considered as likely candidate for aurora observations. They provide unique information about the solar and aurora activities in the first millennium BCE.

# 1. INTRODUCTION

The solar activity has been monitored by telescopic observations of sunspots for more than 400 years (Zolotova & Ponyavin 2015; Vaquero & Vázquez 2009; Owens 2013). Reconstructing the earlier solar activity has been of great interest from the viewpoints of the long-term variation of solar magnetism and its effect on the Earth climate (Haigh 2007; Miyahara 2008; Hathaway 2010; Usoskin 2013b). One common way to reconstruct the past solar activity is the analysis of cosmogenic radioisotopes such as carbon-14 content in tree rings and beryllium-10 in ice cores (Steinhilber et al. 2009).

Records in historical documents provide another way to investigate the solar activity in the pre-telescopic era (Willis et al. 1996; Vaquero & Vázquez 2009; Usoskin et al. 2007; Usoskin 2013b; Hayakawa et al. 2016b; 2016c, submitted). Eddy (1977b, 1980) and Vaquero & Vázquez (2009) even claim that historical documents can offer us data with more precise date than radioactive isotopes.

Chinese official histories (Yau et al. 1988; Yau et al. 1995; Xu et al. 2000; Hayakawa et al. 2015; Kawamura et al. 2016; Hayakawa et al. 2016d, submitted) and Korean official histories (Lee et al. 2004) are especially suitable source documents for the purpose of studying the long-term variation because these sources provide well-formatted records based on continuous observations by professional astronomers (Keimatsu 1976; Hayakawa et al. 2015).

Recently, the past solar activity is also attracting the interest from the viewpoint of extreme space weather (e.g., Tsurutani et al. 2003; Schrijver et al. 2012) particularly after the discoveries of "superflares" in solar-type stars (Maehara et al. 2012; Shibayama et al. 2013; Maehara at al. 2015) as well as the sharp increase of cosmogenic radioisotopes in tree rings around CE775 and CE994 (Miyake et al. 2012; 2013; Mekhaldi et al. 2015), for which one of possible causes are suggested to be large solar flares. The discoveries of "superflares" in solar-type stars posed the question whether such extremely intense solar flares and space weather events can occur in our Sun (e.g., Shibata et al. 2013; Aulanier et al. 2013).

The most intense solar flare throughout the history of telescopic observation is believed to be so-called Carrington event (Carrington 1859; Tsurutani et al. 2003; Cliver & Dietrich 2013). It was associated with worldwide aurora observations by amateur observers, even at low geomagnetic latitudes (Loomis 1859-65; Kimball 1960; Nakazawa et al. 2005; Green & Boardsen 2006; Hayakawa et al. 2016b). A sharp spike of nitrate in polar ice cores has also been found associated with this event (McCracken et al. 2001), although the usage of nitrate as an index of solar flares is debated (Wolff et al. 2012). Searches for the historical records corresponding to the cosmic ray event in CE775 and CE994 (Miyake et al. 2012; 2013) have been carried out by several groups (Usoskin et al. 2013a; Chapman et al. 2015; Stephenson 2015; Hayakawa et al. 2016a; Hayakawa et al. 2016e, submitted).

From the viewpoint of the history of science, it is also interesting to investigate how far back we can trace the solar and space weather events using historical records. Previously, the oldest record was thought to be a Chinese record from BCE 193 (Yau et al. 1995). However, Stephenson & Willis (2002) as well as Stephenson et al. (2004) found a much earlier record from BCE 567 within the Astronomical Diaries from Babylonia (ADB) to be the oldest observation of aurora in the world and the sole aurora observation in ADB.

In this article, we made further careful examinations of ADB in order to search for the potential records of an aurora and hence to obtain insights into the aurora observations and records in pre-Christ era that were not previously well known due to the shortage of historical documents.

## 2. Astronomical Diaries from Babylonia

ADB is a series of Akkadian cuneiform texts inscribed on clay tablets. Most of the tablets were rediscovered in excavations at the site of Babylon in the late 19th century and now preserved in the British Museum, London. Contemporarily, each text of these tablets was titled "regular observation (*naṣāru* (EN.NUN) *šá gi-né-e*)." These records were mostly written and compiled by the families of astronomer-astrologers (*ṭupšar Enūma Anu Enlil*) sponsored by the assembly of Esangil, the temple of god Marduk in Babylon (N 32°33′, E 44°26′) (Mitsuma 2012; Mitsuma 2015). The scholars compiled the diaries from generation to generation, at least from the mid-seventh to the mid-first centuries BC. Toward the mid-third century BC, they fixed the format of the diaries, especially of the so-called "Longer Diaries" or "Standard Diaries."

Each tablet of the standard diaries covers half a year, i.e. six months or seven months if an intercalary month is included, or a third of a year, i.e. four months or five months including an intercalary month. Horizontal rulings divide each tablet into four, five, six or seven sections, according to its coverage. Each section covers a month, and its entries are arranged into subjects in the following order; daily report of the sky, price list of commodities, summary of the positions of the visible planets, level of the Euphrates, and unusual historical event(s).

In this article we mainly examine the daily sky reports of ADB. They use consistent terminology and have a consistent set of criteria for the choice of what should be recorded. Their observations were carried out continuously as the original title of the diary tablet suggests. The astronomer-astrologers inserted a passage "I did not watch (NU PAP)", when they could not make their observations (Sachs & Hunger 1988; Mitsuma 2009).

Importantly, we have only fragments of the original ADB series, so our analysis is limited. Although the earliest known tablet in the ADB dates from BCE 652 and the latest up to BCE 61, we have not unearthed or reconstructed all tablets for every month. We have at most 5-10 % of the estimated original complete ADB series (Stephenson et al. 2004; see also Sachs 1974). Future reconstruction and deciphering of undated clay tablets could improve this situation at least partially.

## 3. METHOD

### 3.1. Target keywords and text survey

Within the ADB, we surveyed records that include descriptions of uncategorized luminous phenomena in the sky, excluding the keywords whose meanings are well known, such as "fall of fire, lightning strike (*miqitti išāti*)", "lightning (*birqu*)", "thunder (*rigim Adad*)", "meteor (*kakkabu rabû*)", "comet (*ṣallammû*)", "halo (TÙR/*tarbaṣu*)", or normal "rainbow (TIR.AN / *manzât*)" shown in the introduction of Sachs & Hunger (1988).

We first surveyed the critical editions of ADB published by Sachs & Hunger (1988, 1989, 1996) and by Hunger & van der Spek (2006). They edited the transliterations of Akkadian texts from all the ADB tablets whose dates became clear by 2006. We then examined the colour photos recently taken by one of the authors (Y Mitsuma) at the British Museum, and copies made by EF Weidner and TG Pinches.

### 3.2. Date conversion

The Babylonian calendar was a lunisolar calendar. One year consists of 12 lunar months (*arḫu*) or 13 lunar months including an intercalary month. Each day starts at sunset, as is the case in the Bible (Parker & Dubberstein 1956, 26; Sachs & Hunger 1988, 15), each month starts when a new crescent moon is observed, and each year starts at the month around the vernal equinox. The Julian dates of the beginnings of Babylonian months covered by the diaries were calculated by Parker & Dubberstein (1956) and by Sachs & Hunger (1988, 1989, 1996), except for those in BCE651, the oldest diary known to us. Note that these conversion tables only cover up to BCE626. We follow the procedure by Sachs & Hunger to convert Babylonian dates in this article. Based on the converted dates, we computed normalized lunar ages according to the algorithm of Kawamura et al. (2016) that employs the interactive data language (IDL) programs in the astronomy user IDL library of NASA/Goddard Space Flight Center (Landsman 1993) and our program developed for numerically determining the minimum of lunar luminosity.

## 4. Result

In total, we found nine candidates of aurora observations recorded in the ADB. Those we found are shown with their ID numbers, keywords, references, dates in Julian calendar, normalized lunar ages, dates in Babylonian calendar, transliterations of Akkadian texts, and English translations. The notation "-n" such as -651 in every entry is the text number of the diaries published by Sachs & Hunger (1988, 1989, 1996) and Hunger & van der Spek (2006). In the section of Babylonian date, "n/n-1 BC" is used to show the Julian equivalent to a Babylonian year. Babylonian months are shown with roman numerals. For the convention used to indicate a part of a diary, see Sachs & Hunger (1988, 36–38).

#1: *manzât* / (very red) rainbow

**Reference:** -651 (BM 32312) Col. iv 20': The text was checked using a recently taken photo of the tablet (Copy: Figure 1).

**Date in Julian Calendar:** ?? ?? BCE 651

**Normalized lunar age:** n/a

**Date in Babylonian Calendar:** BCE 652/651. XII. 28

**Transliteration:** 28 ŠÈG *i ina* KIN.SIG TIR *ma-diš* SA$_5$ *ina* KUR GIB

**Translation:** The 28$^{th}$, a little rain. In the afternoon, a very red rainbow stretched in the east.

#2: *akukūtu* / red glow

**Reference:** -567 (VAT 4956) 'Rev. 10': The text was checked using the copy of the tablet made by EF Weidner, attached as Plate 17 to van der Waerden (1952–1953).

**Date in Julian Calendar:** 12/13 March BCE 567

**Normalized lunar age:** 0.003

**Date in Babylonian Calendar:** BCE 568/567. XI. 29

**Transliteration:** GE$_6$ 29 *a-ku$_6$-ku$_6$-{ku$_6$}-tu$_4$ ina* ŠÚ KUR 2 DA[NNA ....]

**Translation:** Night of the 29th, red glow flared up in the west; 2 double[-hour ....]

#3: *manzât* / rainbow (before sunrise)

**Reference:** -384 (BM 34634) 'Obv. 4': The text was checked using a recently taken photo of the tablet (Copy: Figure 2).

**Date in Julian Calendar:** 8/9 or 9/10 December BCE 385

**Normalized lunar age:** 0.260 or 0.294

**Date in Babylonian Calendar:** BCE 385/384. IX. 8 or 9

**Transliteration:** [....] *la-am* KUR-ḫa TIR.AN Á SI *u* MAR ˹GIB˺ [....]

**Translation:** [....] before (sun)rise, a rainbow stretched in the northwest direction [....]

#4: *dipāru* / torch

**Reference:** -165A (BM 32844) 'Rev.' 10'–11' (according to the reconstruction by Sachs & Hunger 1989, 489)

**Date in Julian Calendar:** 16/17 September BCE 166

**Normalized lunar age:** 0.142

**Date in Babylonian Calendar:** BCE 166/165. VI. 4

**Transliteration:** [4 *ina še-rì*] IZI.GAR TA ULÙ *ana* SI DIB-*ma* UD-DA-*su* [....]

**Translation:** [The 4th, in the morning,] a "torch" crossed (the sky) from the south to the north, and its bright light [....]

#5: *sūmu* / redness

**Reference:** -144 (BM 34609 [+] 34788 + 77617 + 78958) 'Obv. 33'–34': The text was checked with a recently taken photo of the tablet (Copy: Figure 3a, 3b).

**Date in Julian Calendar:** 21/22 September–19/20 October BCE 145

**Normalized lunar age:** n/a

**Date in Babylonian Calendar:** BCE 145/144. VII

**Transliteration:** ITI BI *su-um i-ṣa* [*ina* GI]Š.NIM *u* GIŠ.ŠÚ GAR.GAR-*an*

**Translation:** That month, a little redness was found repeatedly in the east and in the west.

#6: *sūmu* / redness

**Reference:** -143A (BM 34045) 'Flake' 21': The text was checked with the copy of TG Pinches, published by Sachs & Hunger (1996, Plate 190).

**Date in Julian Calendar:** 12/13 July-10/11 August BCE 144

**Normalized lunar age:** n/a

**Date in Babylonian Calendar:** BCE 144/143. IV

**Transliteration:** ITI BI *su-um* [*ina*] GIŠ.NIM *u* GIŠ.ŠÚ GAR.GAR-*an*

**Translation:** That month, redness was found repeatedly in the east and in the west.

#7: *dipāru* / torch

**Reference:** -136B (BM 45745) 'Obv.' 4'-5' (according to the reconstruction by Sachs & Hunger 1996, 182)

**Date in Julian Calendar:** 10/11 or 11/12 November BCE 137

**Normalized lunar age:** 0.710 or 0.743

**Date in Babylonian Calendar:** BCE 137/136. VIII. 20 or 21

**Transliteration:** ˊ20ˋ [.... GE$_6$ 21$^?$ ….] ˊGIMˋ IZI.GAR SAR-*uḫ* UD.DA-[*su*] ˊxˋ GAL IGI-LÁ-*át*

**Translation:** The 20th, [.... Night of the 21st (?) ….] flared up like a torch, [its] bright light was seen.

#8: *manzât* / (red) rainbow

**Reference:** -122A (BM 45998 + 46049) 'Obv.' 11' (according to the reconstruction by Sachs & Hunger 1996, 290)

**Date in Julian Calendar:** 28/29 or 29/30 April BCE 123

**Normalized lunar age:** 0.210 or 0.244

**Date in Babylonian Calendar:** BCE 123/122. II. 6/7

**Transliteration:** [x x] ˊ*ù*$^?$ˋ ŠÚ *šá sin* $^d$TIR.AN.NA *šá* MÚŠ$^{meš}$-*šú* SUD TA SI *ana* Á ULÙ [GIB ....]

**Translation:** [....] and (?) setting of the moon, a rainbow whose shine was red [stretched] from the north to the south side [....]

#9: *dipāru* / torch

**Reference:** -118A (BM 41693) 'Rev. 10': The text was checked using the copy of TG Pinches, published by Sachs & Hunger (1996, Plates 249–251)

**Date in Julian Calendar:** 24/25 October BCE 119

**Normalized lunar age:** 0.754

**Date in Babylonian Calendar:** BCE 119/118. VII. 22

**Transliteration:** [.... SAG] GE$_6$ IZI.GAR TA KUR *ana* MAR SUR-*ma* SUR-*šú ma-gal* BABBAR

**Translation:** [.... beginning] of the night, a torch flashed from east to west, and its flash was very white.

## 5. Discussion

In our survey of the ADB, we found nine records that can be considered as the candidates for aurora observations. They can be categorized using the following keywords: unusual "rainbow (TIR.AN/*manzât*)" (#1, 3, 8), "red glow (*akukūtu*)" (#2), "redness (*sūmu*)" (#5, 6), and "torch (IZI.GAR/*dipāru*)" (#4, 7, 9).

### 5.1. Red glow: *akukūtu*

The record #2 is the well-known record from Stephenson et al. (2004), where the term was introduced and examined philologically and scientifically. The term *akukūtu* means "flame, blaze" or "red glow in the sky (as a rare meteorological phenomenon)" (CAD: I-1, p285). It is the second meaning that appears in this sentence. CAD (*The Assyrian Dictionary of the Oriental Institute of the University of Chicago*) relates this term to aurora as well.

Stephenson et al. (2004) made a positive interpretation of #2 because its last signs - 2 DANNA - "2 double hours" indicate the duration of the phenomenon and there is no other light source that can supply red light for as long as 4 hours. Their conclusion can be reinforced with the flat topography in Babylonia. Babylon is located in the alluvial plain caused by the rivers of Euphrates and Tigris, in an area with no mountains until the Zagros Mountains about 180 km away in the northeast. Its weather is very dry, allowing few trees to grow. This means there is nowhere that a mountain fire or long-lasting light can be caused on the ground.

### 5.2. Unusual rainbow: TIR.AN (*manzât*)

$^{(d)}$TIR.AN(.NA) (*manzât*) of #3, and #8 usually means as "rainbow" or "a name of star" or sometimes as "halo" with a subsequent description such as "surrounding the sun/moon" (CAD, X-1, pp230-232). #1 shows TIR. It seems to be an abbreviated form of TIR.AN and is followed by the words *ma-diš* SA$_5$ "very red," as is suggested by the comment of Sachs & Hunger (1988, p46) to #1. *manzât* was used with the verb GIB (*parāku*) "stretch" in #1 and #3, and the same verb may be restored in #8. These "rainbows" appear "stretched" in the sky with a red colour (#1, #8), or at night (#3, #8). Since these appearances do not match the nature of normal rainbows, *manzât* in #1, #3, and #8 seem to have been something different.

The unusual "rainbow" or TIR.AN in #3 was observed "before sunrise." The lunar age of this event is about 0.3 with the moon at about the waxing quarter, and the moon had already set at this time "before sunrise." Therefore, moonbow or any other moon-related atmospheric optical phenomenon cannot explain this event. Considering that this event was seen in the northern direction, which is usual for low-latitude auroras, it is reasonable to leave this record as the candidate for an aurora observation.

The term ᵈTIR.AN.NA (*manzât*) in #8 is recorded after the "setting of the moon". The exact observational date is broken in the original tablet. However, this record is located between the record for BCE 123/122. II. 5 and that for BCE 123/122. II. 8 and thus can be located on II. 6/7, namely on 28/29 or 29/30 April BCE 123. Its moon phase is 0.227 or 0.261 and thus the moon was approximately in its waxing quarter. This means "after moonset" was equivalent to just after midnight. This red "rainbow" may be regarded as an auroral arc elongated "from north to south" across the sky.

The unusual "rainbow" or TIR in #1 was also red in colour. Unfortunately the Julian date of this record (BCE 652/651. XII. 28) was not calculated by Sachs & Hunger (1988) nor by Parker & Dubberstein (1956). Despite its abnormality, this record is very unlikely to represent aurora due to the observation time being "in the afternoon (KIN.SIG)". Red rainbows can be observed especially at the time of twilight when red-coloured light from the sun gets refracted by waterdrops. The "rain" immediately before this phenomenon might have supplied the water-drops to cause rainbows in the quite dry weather in Babylonia.

Nevertheless we should note two historical reports for luminescence phenomena observed in the day time in a huge magnetic storm, namely in the Carrington event. Loomis (1860b) cited a letter from Lieut. N. Home to describe aurora observation at Halifax on 28 August 1859. At 17:00, "a long narrow belt of cloud from E. to W. having a peculiar orange-white appearance" was seen, and at 20:00 "this cloud suddenly became luminous at its eastern extremity". This cloud is unlikely to have been aurora in the day time, but we cannot rule out the possibility that it was aurora.

Meanwhile, "unusual rainbow" can sometimes be related to bow-shaped auroras worldwide. A Norwegian drawing for aurora on 26 November 1710 was drawn like a rainbow in Ramus (1745) and Chinese historians reported some auroras as "white/unusual rainbow" in their official histories as highlighted by Hayakawa et al. (2016a).

### 5.3. Redness: *sūmu*

The term *sūmu* used in #5 and #6 means "redness, red glow" or "red spot". While the term in the latter meaning is used for red spots on the body, the term in the former meaning is used in astronomical contexts (CAD: XV, pp381−383). The *sūmu* of #5 and #6 were probably celestial events, although they appear not in reports of celestial events but in reports of historical events. Those observations of "redness" might have been considered as omens for historical events and recorded with them. Another celestial event, for example, lightning strike (*miqitti išāti*), is often inserted into the historical parts of the diaries and seems to have an ominous significance (Pirngruber 2013).

These terms of "redness" are used almost in the same expression with the signs GAR.GAR-an, which represents the Akkadian verb *ittaškan*, Ntn, i.e. passive habitative-iterative (Caplice & Snell 2002: pp.51-52), preterite of *šakānu* "to place." This indicates that the "redness" of #5 and #6 appeared repeatedly.

These philological analyses show us that "redness" appeared repeatedly in the same month in both cases. This phrase may describe low-latitude auroras whose movement is quite slow and which thus seem to remain in the same place for a long time. During the Halloween storm (October 29-31, 2003), aurora observations were made for 3 successive nights (Shiokawa et al., 2005). Alternatively, this phrase is likely to describe a stable red auroral (SAR) arc that can be seen as red, faint lights. Data from optical imaging on board Dynamics Explorer 1 show a SAR arc lasting for 28 hours (Craven et al., 1982). In fact, auroras can appear repeatedly for several days when complex active regions on the sun continue to launch multiple coronal mass ejections. At the Carrington event, the strong magnetic storms brought a cluster of auroras to the earth from 1859/08/28 to 09/04 (Loomis 1859-65; Kimball 1960; Green & Boardsen 2005; Hayakawa et al. 2016b). In September 1770, a series of aurora observation were made in East Asia for at least three days straight (Willis et al. 1996; Kawamura et al. 2016). These records (both #5 & #6) suggest the possibility of strong solar activity in BCE 145-144. In summary, although it is slightly mysterious that they appeared in the east and west, these records are not contradictory with aurora records.

**5.4. Torches: IZI.GAR (*dipāru*)**

This term "IZI.GAR (*dipāru*)" from records #4, #7, and #9 means "torch" (CAD: III, pp156–157). The phrase "*ḫakukūtu*, which is like *dipāru*" appears once in a divination text (Virolleaud 1911–1912, no.107:3).

Record #4 involves a verb "DIB (*etēqu*)". This is a motion verb meaning "to pass/go along" (CAD: IV, p384; Sachs & Hunger 1988, p30). This means that the "torch" in #4 moved from south to north and is probably a meteor or a fireball. If we interpret this record as aurora, we could relate this with expansion and contraction of the aurora oval.

Phenomenon of record #9 has an extension from east to west. The problem is to identify if this is with motion. Its verb "SUR (*ṣarāḫu*)" means "to light up," "to flare up," or "to display a sudden luminosity" (CAD: XVI, p100). This verb itself does not provide an answer to our question whether this recorded phenomenon appeared without motion. There is a possibility that this is aurora. Bright aurora looks white or greenish white because it consists of emissions from O (green colour) and from $N_2$ and $N_{2+}$ (red and blue colours). After auroral breakups, bright aurora expands toward the north and west. The leading edge of the bright aurora is called a westward traveling surge. The westward traveling surge moves westward, but it is unclear if such an elongated structure is expressed as a torch. Thus, this reminds us of a fireball.

As for record #7, its subject is lost. This makes any detailed discussion difficult. However, something like "a torch" appeared with "UD.DA (*ṣētu*, bright light)." The word *ṣētu* frequently refers the light of the sun or the moon (CAD: XVI, pp150-153). Therefore, strong brightness, at least, seems to be associated with this recorded phenomenon. According to the date written in the clay tablet, its date is 10/11 or 11/12 November BCE 137. In case if it is observed at daytime, we can hardly regard this event as aurora. In case if it is observed at night on 11/12, we have still a little possibility to regard this event as aurora. We have examples for low-latitude aurora, which becomes partially brighter (Shiokawa et al. 2005). Thus, we cannot exclude the possibility of aurora for this record.

### 5.5. Likeliness of every record

Based on the analysis above, we can rate these aurora-like records as follows (Table 1). #2 has no reason not to be related with aurora as Stephenson et al. (2004) stated. The same can be stated for the "redness" in #5 and #6. They appeared repeatedly and remind us of clustering aurora caused by huge geomagnetic storms like those at the Carrington event. The unusual "rainbows" in #3 and #8 are not unlike aurora, since these rainbows are observed at night. Despite the same terminology, #1 is unlikely to be aurora due to its daytime observation. #4 and #9 seem like fireballs or meteors from the descriptions and we do not have any positive intention to relate them with aurora. #7 could be aurora as the term *dipāru* can be used for description of *ḫakukūtu*, a variant of *akukūtu*, which can mean aurora, though its subject is lost and we cannot make a clear conclusion on this.

Korte & Stolze (2012) estimated the location of the aurora zone for past 10,000 years. During the period spanning from BCE 652 to BCE 61, Babylon was situated near the center of the two boundaries where aurora was visible on the horizon at the geomagnetic activity levels Kp=4 and Kp=9. According to the statistical study by Remick & Love (2006), the mean wait times between successive events are 7.12, 16.55, 42.22, and 121.40 days for Kp ≥ 5, 6, 7, and 8, respectively. Thus, it is quite likely that aurora was frequently visible at Babylon somewhere between the horizon and zenith.

**5.6. Comparison with long-term solar activity**

In order to map the aurora-like records shown above with long-term solar activity, we compared the dates of these records with proxy-based solar activity levels. Figure 4 shows a comparison of the aurora-like records with the solar activity level anomaly reconstructed using multiple proxies of radioactive isotopes by Steinhilber et al. (2009) at a resolution of 20-30 years. However, it should also be noted that this comparison between characteristics of the solar activity at different time-scales is hard and complex, as on one hand auroras occur at daily scale and on the other hand the reconstruction of solar activity by Steinhilber et al. (2009) has a decadal time-resolution. (see Vaquero et al. (2002) for the similar problems in case of naked-eye observations of sunspots).

This figure shows us that most of the aurora-like records, except for #3 in BC 385 in Solar Minimum (Eddy 1977a, b; Usoskin et al. 2007; Nagaya et al. 2012), are located in phases of high solar activity. The cases #5-#7 are clustered in BCE 140s-130s and #8-#9 are around BCE 120. These peaks are comparable with contemporary aurora records from other regions. For the former peak, red vapour was recorded in BCE 139 in Chinese official history *Hànshū* (Yau et al. 1995), and reports of sky fire in BCE 147 and night sun in BCE 134 in western classics (Stothers 1979). For the latter peak, we have reports of milk rain in BCE 124 and BCE 118 in western classics (Stothers 1979). Although we do not have exact simultaneous observation of aurora, these clustering records support the peaks at both BCE 140s-130s and around BCE 120 shown in the ADB.

**6. Conclusion**

We surveyed aurora-like records in ADB spanning from BCE 652 to BCE 61. We found 9 records of aurora-like phenomena including one mentioned in Stephenson et al. (2004) and examined them to evaluate 5 likely, 3 unlikely and 1 possible aurora event. The main characteristics of these events are provided in this article. Our result is quite consistent with the long-term solar activity level, although comparison between characteristics of the solar activity at different time-scales has difficulties as stated above. This article rewrites the history of aurora observation in this early period with dates to provide clear insights for the era before Christ. This article also examines proxies at a high resolution in this early period in order to consider long-term solar activity and rare extreme space weather events. A lot of unpublished clay tablets of the ADB are preserved in the British Museum. Although most of them are fragmentary, study of those tablets will provide us further astronomical information from the mid-seventh to the mid-first centuries BCE, possibly including further records of aurora-like phenomena.

## Abbreviations

BM: Tablets in the collections of the British Museum

VAT: Tablets in the collections of the Staatliche Museen, Berlin

## Competing interests

There are no competing interests.

## Authors' contributions

This research was performed with the cooperation of authors as follows: HH and YM made historical and philological contributions. ADK, YE, and HT made contributions on scientific interpretations and analyses. HM compared and discussed records with long-term solar activity. HI supervised this study. All authors read and approved the final manuscript.

## Acknowledgement

We thank the Trustees of the British Museum for allowing the study and photography of the tablets BM 32312, BM 34609 (+) 34788 + 77617 + 78958, and BM 34634.

We acknowledge support from the Center for the Promotion of Integrated Sciences (CPIS) of SOKENDAI as well as Kyoto University's Supporting Program for Interaction-based Initiative Team Studies "Integrated study on human in space" (PI: H. Isobe), the Interdisciplinary Research Idea contest 2014 by the Center of Promotion Interdisciplinary Education and Research, the "UCHUGAKU" project of the Unit of Synergetic Studies for Space, and the Exploratory Research Projects of the Research Institute of Sustainable Humanosphere, Kyoto University. This work was also supported by Grant-in-Aid from the Ministry of Education, Culture, Sports, Science and Technology of Japan, Grant Numbers 15H05816 (PI: S. Yoden), 26870111 (PI: Y. Mitsuma), and 15H05815 (PI: Y. Miyoshi).

**Figures**

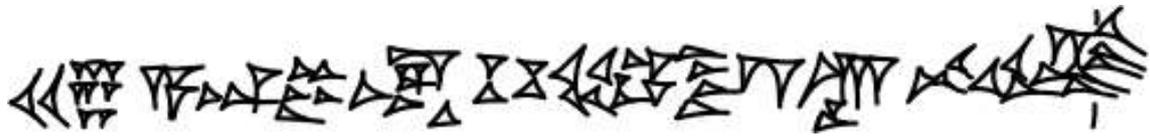

**Figure 1:** A copy of #1, -651 Col. iv 20', by Yasuyuki Mitsuma

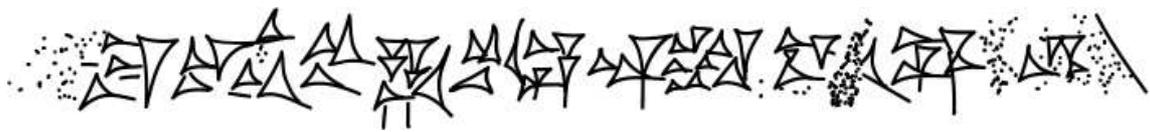

**Figure 2:** A copy of #3, -384 'Obv. 4', by Yasuyuki Mitsuma

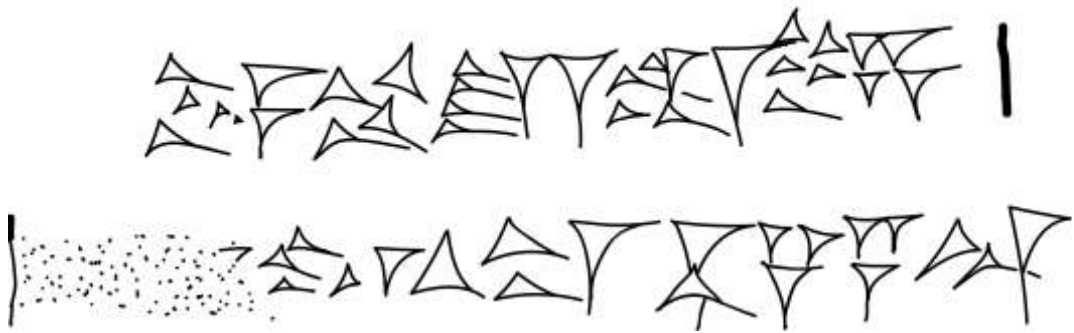

**Figure 3a (upper part):** A copy of #5, -144 'Obv. 33' right side, by Yasuyuki Mitsuma
**Figure 3b (lower part):** A copy of #5, -144 'Obv. 34' left side, by Yasuyuki Mitsuma

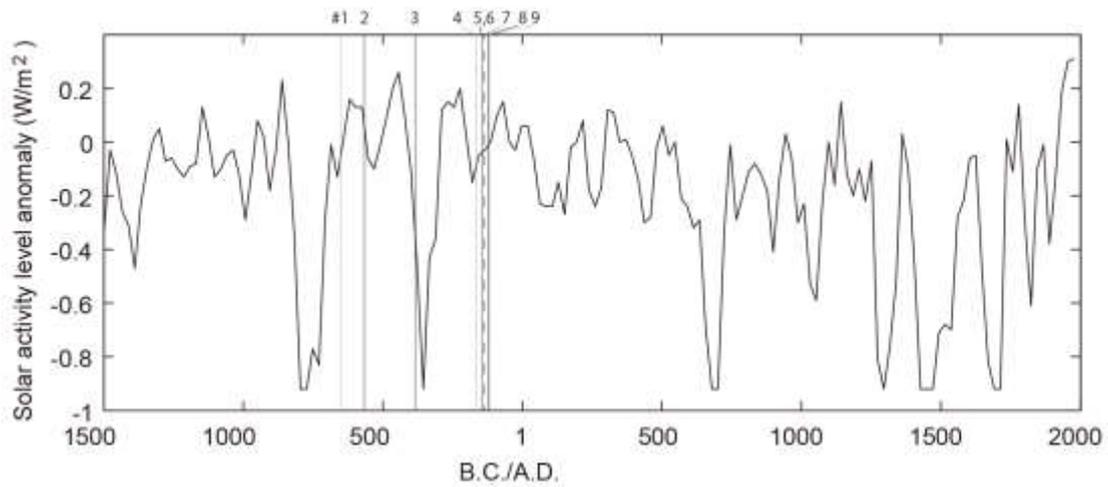

**Figure 4:** Aurora-like records compared with total solar irradiance anomaly

Black Curve: Total solar irradiance anomaly from Steinhilber et al. (2009)

Black vertical lines: Likely aurora records from the ADB

Gray vertical lines: Unlikely aurora records from the ADB

Black dashed vertical lines: Possible aurora records from the ADB

| ID | year (BCE) | month | date | keyword | direction | likeliness |
|---|---|---|---|---|---|---|
| #1 | 651 | ?? | ?? | *manzât* | E | unlikely |
| #2 | 567 | 3 | 12/13 | *akukūtu* | W | likely |
| #3 | 385 | 12 | 8/9 or 9/10 | *manzât* | NW | likely |
| #4 | 166 | 9 | 16/17 | *dipāru* | S to N | unlikely |
| #5 | 145 | 9/10 | | *sūmu* | E and W | likely |
| #6 | 144 | 7/8 | | *sūmu* | E and W | likely |
| #7 | 137 | 11 | 10/11 or 11/12 | *dipāru* | | possible |
| #8 | 123 | 4 | 28/29 or 29/30 | *manzât* | N to S | likely |
| #9 | 119 | 10 | 24/25 | *dipāru* | E to W | unlikely |

**Table 1:** Summary table for records of aurora-like phenomena in ADB